\input{epsf}
\epsfverbosetrue
\documentstyle[prl,aps,multicol,epsf]{revtex}
\renewcommand{\narrowtext}{\begin{multicols}{2}
\global\columnwidth20.5pc}
\renewcommand{\widetext}{\end{multicols}
\global\columnwidth42.5pc}
\begin{document}
\draft

\title{Obervation of non-linear stationary spin waves in superfluid $^3$He-B}
\author{A.~S.~Chen, Yu.~M.~Bunkov,  H.~Godfrin, R.~Schanen,
F.~Scheffler.}
\address{ Centre de Recherches sur les Tr\`es Basses
Temp\'eratures, \\Centre National de la Recherche Scientifique,\\
BP 166, 38042 Grenoble, Cedex 9, France.}
\date{\today} \maketitle
\begin{abstract}

Due to its broken spin and orbit rotation symmetries, superfluid $^3$He
plays a unique role for testing rotational quantum properties on a
macroscopic scale.
In this system the orbital momentum forms textures that provide an
effective potential well for the creation of stationary spin waves.
In the limit of the lowest temperatures presently attainable, we observe by
NMR techniques a profound change in the spin dynamics.
The NMR line shape becomes asymmetric, strongly hysteretic and displays
substantial frequency shifts.
This behavior, quantitatively described by an anharmonic oscillator model,
indicates that the parameters of the potential well depend on the spin
waves amplitude, and therefore that the orbital motion is not damped in
this new regime, not considered by the standard Leggett-Takagi theory.
This regime of non-linear stationary spin waves is shown to give rise to
the pulsed NMR "Persistent Signals" reported recently.
\end{abstract}

\narrowtext

The ground state of superfluid $^3$He is formed by Cooper pairs of $^3$He
atoms in a triplet state with Spin and Orbital momentum quantum numbers
equal to unity.
In consequence superfluid $^3$He  shows not only the usual superfluid
properties arising from the broken gauge symmetry, but also displays
macroscopic quantum rotation phenomena related to the broken spin and orbit
rotation symmetries.
The spin concerned is that of the nuclear rotation and is thus associated
with a nuclear magnetic moment.
The orbital motion consists of the mutual orbiting of two neutral $^3$He
atoms and is thus not associated with a magnetic moment.
These two motions are not independent, since the dipole-dipole interaction
of two nuclear magnetic moments depends on their relative orientation.

In the $^3$He-B  phase, at zero field, the magnetization and the average
orbital momentum are equal to zero.
Applying an external magnetic field induces a magnetization $\vec M$ and,
via the spin-orbit symmetry of the wave function, an orbital momentum $\vec
L$.
This symmetry is usually characterized by the rotation matrix $\hat R$ of
angle $\Theta$ around an axis $\vec n$.
The orbital momentum is therefore $\vec L=\hat R(\vec n, \Theta)\vec M$.

The early investigations of superfluid $^3$He have been performed at
relatively high temperatures of order 0.4 T$_C$ and above.
In these conditions, first, the orbital motion is blocked due to an
effective interaction with quasiparticles.
Consequently, the NMR of the spin system considered up to now only involves
quasi-stationary orbital dynamics.
Second, there is an effective magnetic interaction between the superfluid
and normal components of $^3$He, which is characterized by an effective
quasiparticle lifetime $\tau$.
One has to distinguish two different regimes of spin dynamics, hydrodynamic
and non hydrodynamic, for $\omega \tau$ smaller or larger than unity (here
$\omega$ is the NMR frequency).
The equations for the spin precession derived by Leggett and Takagi
(L-T)\cite{LT}, written down for hydrodynamic conditions, describe well the
high temperature experimental results.
The superfluid component is responsible for spin supercurrents, the order
parameter stiffness and the dipole frequency shift, while the normal
component determines magnetic relaxation and spin diffusion (a review is
given in ref.\cite{rev}).

It is usually believed that these equations remain valid in
non-hydrodynamic conditions, with some sort of renormalisation of the
relaxation parameters.
This renormalisation was tested in NMR experiments in the temperature
region above about 0.4 T$_C$ and found to be in good agreement with the
theoretical predictions\cite{BE}.
The effective value of the magnetic interaction decreases very rapidly in
the non-hydrodynamic region, as $1/(\omega \tau)$.

In the present work, performed at much lower temperatures, we are able to
reach a new condition where the spin and orbit dynamics of the superfluid
component are no longer damped by the interaction with the normal
component, where we can study true superfluid properties of rotation
quantum dynamics.
We describe in this article the observation of a new non-linear phenomenon
at temperatures of about 0.2 T$_C$, now accessible to experimental
investigation.

Let us first remind the properties of spin dynamics at temperatures above
0.4 T$_C$.
First, the equilibrium local orientation of the orbital momentum $\vec L$
with respect to the external magnetic field supplies an additional
potential for NMR and leads to a NMR frequency shift.
Usually the orbital momentum orientation is determined by boundary
conditions at the walls of the cell, magnetic field anisotropy and bending
energy.
Since the angle $\Theta $ is fixed at about $104^\circ$ by dipole-dipole
interactions, the relative orientation of $\vec M$ is then determined by
$\vec n$ which can form a complicate texture.
The local NMR frequency at relatively high field is:
$$ \omega^2= (\gamma H)^2 + n_\perp ^2 \Omega_D ^2 \eqno(1)$$
where $\Omega_D$ characterizes the dipole-dipole interaction and $n_\perp$
is the component of $\vec n$ normal to the magnetic field.
The distribution of the orientation of $\vec L$ gives rise to a broadening
of the NMR line by about ${\Omega_D^2 \over \gamma H}$, which can be on the
order of tens of kHz.

Second, owing to the stiffness of the order parameter, the spin system
exhibits non-local resonant modes - Stationary Spin-Waves modes (SSW)
trapped by a texture.
These modes were observed in the pioneering work by Osheroff \cite{osh}.
Later SSW were used for studying spin diffusion \cite{2}, textures and
vortices \cite{3}.
The SSW can be created in a potential well determined by the walls of the
cell and the field gradient \cite{2}, by the wall and a texture \cite{osh}
or only by a texture, for example the "Flared-out" texture \cite{3}.
The equation of motion for the transverse component of the magnetization
$M^+$ in the texture potential well $\beta _L(r)$ has the form of the
Schr\"odinger equation \cite{osh,3}:
$$ (\omega - \omega (z))M^+ =
 { \Omega_D^2\over {2 \omega (z)}}( { 2\over 5}
\sin^2\beta _L(r) -
{ 24\over 65} \xi _d^2 \nabla ^2)M^+.\eqno(2)$$
where $\beta _L$ is the local deflection of the orbital momentum from the
direction of magnetic field  and $\xi_d$ the dipole coherent length.
The SSW modes have been studied at relatively high temperatures, above 0.4
T$_C$,  and found to be in a good agreement with theory.

The new Grenoble refrigerator allows us to cool $^3$He to significantly
lower temperatures.
Owing to the exponential dependence of the number of quasiparticles on
temperature, we are able to decrease their density by many orders of
magnitude, thus suppressing the magnetic and orbit friction.
In this article we report the observation of a dramatic change in the spin
and orbit dynamics, observed in a  CW-NMR experiment in this ultra-low
temperature region.

The experiments were performed in a cylindrical experimental cell
of diameter 6 mm and height 5 mm, at
pressures ranging from 0 to 6 bars. The magnetic field, corresponds to a
NMR frequency of
about 500 kHz and was directed parallel to the axis of the cell. In
addition, we were able to
change the magnetic field gradient in order to determine the localisation
of NMR signals within
the cell. In these conditions a "flared out" texture should be formed.
This is
shown schematically in the inset of Fig.2:
at the top and the bottom of the cell, the orbital momentum is oriented
parallel to the
field, while the vertical walls orient $\vec L$ perpendicular to the
field. This textural configuration gives a region of minima of energy for
NMR of SSW near the top and the bottom of the cell.

We observed SSW signals at temperatures down to 0.22 T$_C$, i.e.
significantly lower than in previous investigations.
 We report here two new features:

a) we observe an enormous number of SSW new modes,

b) the SSW modes exhibit strong non-linear behavior.

\begin{figure}[htb]
\centerline{\epsfxsize=9 cm \epsfbox{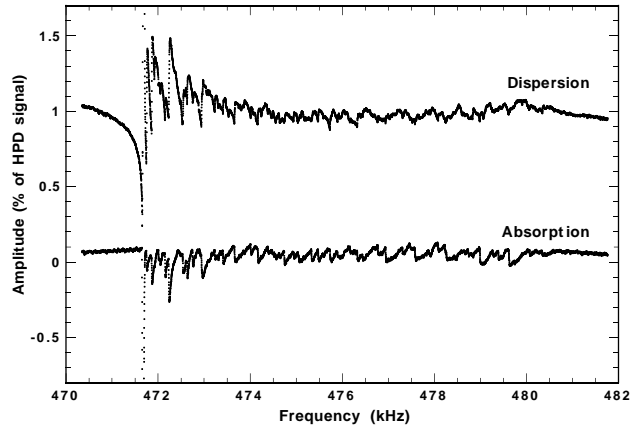}}
\bigskip
\caption{The whole CW NMR line at 0.22 T$_C$. Excitation level:
H$_1$=10$^{-5}$ Tesla, magnetic gradient: 5 10$^{-4}$ Tesla.m$^{-1}$.
Nearly 100 SSW modes are observed.}
\label{fig1}
\end{figure}

As shown in Fig.1 we observe at this temperature an enormous number of NMR
peaks that cover practically all the NMR line.
By changing the magnetic field gradient, we first determine the positions
within the cell that generate these SSW modes.
We performed a calibration in normal $^3$He and then, at 0.22 T$_C$, we
studied the frequency vs. magnetic field gradient dependence of each peak
in the limit of small excitation level.
This allows us to show that the two largest peaks correspond to the top and
to the bottom of the cell (as was expected in this textural configuration,
and confirmed by the same study on SSW at 0.44 T$_C$), with dipole shifts
from the Larmor frequency on the order of 150 Hz.
Finally all the peaks correspond either to the fundamental state at
different places within the cell, or to harmonics.

We then fixed the magnetic field gradient at a value of 5 10$^{-4}$
Tesla.m$^{-1}$ and studied the behavior of several low-frequency-shift SSW
modes, when changing the excitation level and the frequency sweep
direction.
In Fig.2 we show the signals of a few low-frequency-shift SSW modes
(absorption and modulus) at 0.44 T$_C$ and 0.22 T$_C$, for different
excitation levels H$_1$
(note that our sweep is given in units of frequency, which is recalculated
from our real field sweep. Sweeping the field up corresponds therefore to
sweeping the frequency down.).
Even if these signals are radiated by the same SSW modes at both
temperatures, their properties are very different:
at low temperatures, the broadening of the signals does not correspond to a
relaxation process, but to a non-linear frequency shift with excitation.
Let us look at signals for a relatively large rf field (H$_1$=1.8 10$^{-6}$
Tesla).
If we sweep the field up (frequency down), the low-temperature-signals
appear at the same frequency as at high temperature, but they grow rapidly
even well below the high temperature resonance frequency. At some critical
point, the signal disappears. In the sweep back we observe a strong
hysteresis since the signal appears only around the high temperature
resonance frequency.
This behavior is typical of a non-linear oscillator, where the frequency
strongly depends on the excitation level. The bigger the excitation, the
bigger frequency shift that can be achieved.
\begin{figure}[htb]
\centerline{\epsfxsize= 7 cm \epsfbox{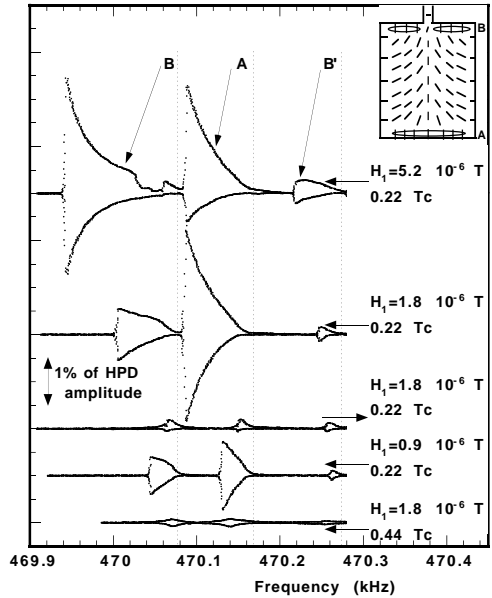}}
\bigskip
\caption{ CW NMR absorption (negative signals) and NMR amplitudes (that
correspond to $M^+$, positive signals) for different temperatures and
amplitudes of rf field, H$_1$.
The arrows indicate the direction of the frequency sweep.
The inset shows the $\vec L$ texture and the regions (labeled A and B)
radiating the SSW signals  A, B, and the higher harmonics, like B'.}
\label{fig2}
\end{figure}

Following Landau and Lifchitz \cite{LL}, we consider an anharmonic
oscillator with a third order non-linearity:
$$
\ddot{M^+}+2\lambda\dot{M^+}+\omega_0^2 M^+= F cos(\omega t)-\alpha
{M^+}^{2}-\beta {M^+}^3
 \eqno(3)$$
The amplitude of the transverse component of the magnetization, $M^+ =
Acos(\omega t)$, is found by solving the equation
$$
4\omega_0 ^2 A^2((\omega -\omega_0-\kappa A^2)^2+\lambda^2)=F^2,
\eqno(4)$$
where $\kappa$ describes the dynamic frequency shift.
Equation (4) is of order six in $A$, and we are only interested in real and
positive solutions.
The coefficients of equation (4) are obtained by
fitting our experimental data.
Fig.3 shows the comparison between the theoretical curve and the
experimental signal amplitude from the top of the cell for a rf field level
of 1.8 10$^{-6}$ Tesla. When sweeping the frequency $\omega$ down, the
amplitude of the transverse magnetization follows the part $1-2$ and then
drops down to
$zero$. When the frequency increases, it stays equal to $zero$ until point
$3$ and
jumps to point $4$, thus showing hysteretical behavior.
We find that $\lambda$=0.001 kHz is the damping factor giving the width of
our resonant curve, while the anharmonic coefficient ($\kappa$=-1700
kHz.A$^{-2}$) relates the frequency to the amplitude of excitation. Here
the amplitude A was measured in units of the full HPD signal (Homogeneously
Precessing Domain: in all the sample, the magnetization is deflected by
$104 ^\circ $ and precesses homogeneously).
For 0.44 T$_C$ we find $\lambda$=0.007 kHz, while the anharmonic
coefficient is negligible.
\begin{figure}[htb]
\centerline{\epsfxsize=6 cm \epsfbox{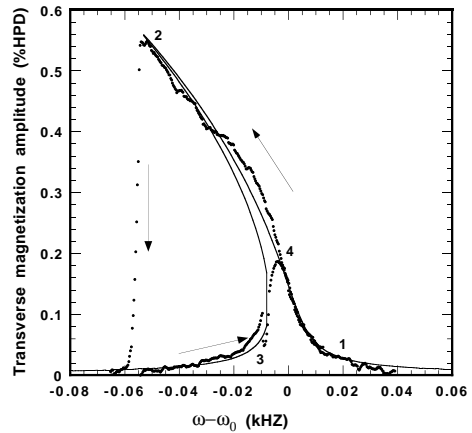}}
\bigskip
\caption{A non-linear SSW signal (points) and its fit by the non-linear
oscillator equation (line).}
\label{fig3}
\end{figure}

This anharmonic oscillator model corresponds to the fact that the potential
well radiating the SSW becomes anharmonic when the transverse magnetization
increases.
This effect can be quantitavely described by incorporating a feedback
correction into the potential term, $\sin^2\beta_L$, of the Schr\"odinger
equation (2): $\beta_L$ should no longer depends only on r, but also on
$M^+$.
Indeed, the non-linear SSW are obtained by pumping the energy into the spin
wave modes localized in the potential well formed by the orbital texture.
When the intensity of a given spin wave increases, it influences the
orbital degrees of freedom.
As a result, the potential well should be modified and change the
eigen-frequency of the spin-wave mode, which in turn affects the orbital
texture.
The frequency of this self-consistent precessing state thus depends on the
excitation level and decreases with the magnetization deflection.
Qualitatively, when the precessing magnetization  is deflected, the texture
becomes shallower, causing a decrease of the spin waves frequency shift.

The signal from the bottom of the cell, labeled A in Fig.2, has a more
complicate behavior. Its amplitude increases faster than in the model
considered here.
Nevertheless, the main features are similar. Possibly, the difference
originates from the spatial extension of the SSW mode at high levels of
excitation.
\begin{figure}[htb]
\centerline{\epsfxsize= 8 cm \epsfbox{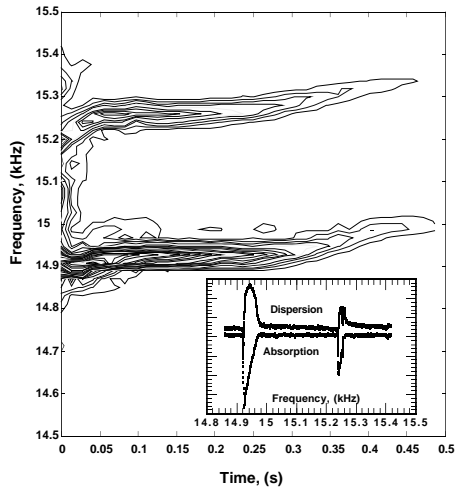}}
\bigskip
\caption{The time dependent spectrum of Persistent Signals observed after a
NMR pulse. Inset: the two main CW NMR signals in the same conditions. (the
base frequency of the pulse NMR spectrometer (467 kHz) was subtracted from
both signals)}
\label{fig4}
\end{figure}

Finally, we investigated the relation between the excitations reported here
and Persistent Signals (PS).
A Persistent Signal  is a small but
extremely long lived induction decay signal that was observed in previous
pulsed NMR experiments\cite{Lan2}.
Its frequency increases with time\cite{Z} (which is opposite to
conventional HPD!), and it is radiated by
a texture\cite{Y}.
We made pulsed and CW NMR, in the same low temperature conditions.
In CW NMR, we observed the non-linear SSW described before, while in pulsed
NMR, we observed Persistent Signals.
We show in Fig.4 the time dependent spectrum of the signal after an NMR pulse.
We find two PS whose frequencies increase with time, while their amplitudes
decrease.
The two PS appear at the same frequencies, and they have the same width and
frequency vs. amplitude dependence as the two main non-linear CW NMR
signals.
So, they seem to be radiated from the same texture by the same mechanism.
This identification is particularly informative on the nature of the PS and
suggests a physical scenario for their creation.
At low temperatures, the magnetization precession with high
deflection angles is very unstable \cite{BGT}.
So, after the NMR pulse, the deflected magnetization decays quickly to the
spin waves modes.
However, since the orbital momentum has also some kind of flexibility, the
PS signal arises as a combined solution for the magnetization and texture
motion.
This explanation is consistent with the feedback mechanism for SSW in a
"flared out" texture proposed earlier to explain some of the PS
properties\cite{X}.
In principle, it should be possible to develop a quantitative theory for PS
as well as non-linear SSW by numerical simulations.
These would be extremely helpful for guiding future experimental work.

In conclusion, we have shown that the spin dynamics in superfluid $^3$He
are profondly modified in the limit of ultra-low temperatures.
In this new regime the stationary spin waves become extremely non-linear,
displaying the typical behavior of a third order anharmonic oscillator.
This effect can be ascribed to a softening of the texture potential for the
spin waves, and modeled by a non-linear Schr\"odinger equation which
governs the transverse magnetization dynamics.
We also show that the origin of the Persistent Signals observed in pulsed
NMR experiments is directly related to the non-linear dynamics of the
system.
Our new qualitative and quantitative results call for an extension of the
"Standard Model of $^3$He" given by Leggett-Takagi in order to describe the
ultra-low temperature regime, where the orbital motion plays an important
role in the spin dynamics.
Finally, we would like to mention that spin waves in superfluid $^3$He-B
can serve as an interesting experimental model for Particle Physics systems
described by the non-linear Schr\"odinger equation.

We are grateful to  I.~A.~Fomin, A.~J.~Leggett,  G.~R.~Pickett,
O.~D.~Timofeevskaya
and G.~E.~Volovik for many stimulating discussions.

\widetext

\end{document}